\documentstyle[epsfig]{aipproc}
\begin{document}
\title{Detecting Eccentric Globular Cluster Binaries with LISA}
\author{M. Benacquista}
\address{Montana State University-Billings, Billings, MT 59101}
\maketitle
\begin{abstract}
The energy carried in the gravitational wave signal from an eccentric binary is spread across several harmonics of the orbital frequency. The inclusion of the harmonics in the analysis of the gravitational wave signal increases the signal-to-noise ratio of the detected signal for binaries whose fundamental frequency is below the galactic confusion-limited noise cut-off. This can allow for an improved angular resolution for sources whose orbital period is greater than 2000 s. Globular cluster sources includ possible binary black holes and neutron stars which may have high eccentricities. Cluster dynamics may also enhance the eccentricities of double white dwarf binaries and white dwarf-neutron star binaries over the galactic sources. Preliminary results of the expected signal-to-noise ratio for selected globular cluster binaries are presented.
\end{abstract}
The space-based gravitational radiation detector LISA will be sensitive to frequencies in the millihertz range. A number of interesting astrophyisical phenomena will be likely sources in this frequency band, many of interest in cosmology and as fundamental tests of relativity. LISA will also be of use as an astronomical tool for observing compact binary systems in the galaxy and globular cluster system. In the globular cluster system, LISA can test dynamical evolution nodels by detecting binary populations which are predicted by simulations but which are exceedingly difficult to observe in electromagnetic radiation. One such class of objects is a population of binary black holes with separations between 3 and 30 $R_{\sun}$ and a thermal distribution of eccentrities \cite{spz,bpzr}. A Monte Carlo analysis of the detection of such signals indicate that eccentricities above 0.8 can extend the detectable systems to orbital periods on the order of 1 day.

The gravitational wave signal from an eccentric binary is given by the two polarizations \cite{Pierro}:
\begin{eqnarray}
h_{\times} &=& \frac{\cos{\vartheta}}{\sqrt{2}}\sum_{n=1}^{\infty}{[2h_{xy}^{(n)}\sin{(2\pi nft)}\cos{2\varphi}-h_{x-y}^{(n)}\cos{(2\pi nft)}\sin{2\varphi}]}\\
h_+ &=& \frac{1}{2\sqrt{2}}\sum_{n=1}^{\infty}{[(1+\cos^2{\vartheta})(2h_{xy}^{(n)}\sin{(2\pi nft)}\sin{2\varphi}+h_{x-y}^{(n)}\cos{(2\pi nft)}\cos{2\varphi})}\nonumber\\
& & ~~~~~~~~~~~~-(1-\cos^2{\vartheta})h_{x+y}^{(n)}\cos{(2\pi nft)}]
\end{eqnarray}
in a coordinate system centered on the binary with the x-axis along the semi-major axis and the z-axis along the angular momentum vector of the system, with the orbital frequency of the binary given by $f$. The angles $\vartheta$ and $\varphi$ are the usual polar coordinates giving the direction to the detector. The metric components, $h_{xy}^{(n)}$ and $h_{x\pm y}^{(n)}$, share a common amplitude which depends upon the masses and the orbital frequency. In general, they are combinations of Bessel functions.

The signal detected by LISA is modified by the motion of LISA in its orbit about the sun. To describe the detected signal, we define several unit vectors. Let the orientation of the arms of the constellation of LISA spacecraft be given by the three vectors, $\ell_1, \ell_2, \ell_3$. The direction from the sun to the source is $\hat{n}$. The orientation of the binary orbit is $\hat{L}$ (which points along the angular momentum), $\hat{a}$ (which points along the semi-major axis), and $\hat{b} = \hat{L} \times \hat{a}$. The gravitational wave propagates along $-\hat{n}$, and we define the polarization axes $\hat{p} = {\hat{n}\times\hat{L}}/{\vert\hat{n}\times\hat{L}\vert}$ and $\hat{q} = -\hat{n}\times\hat{p}$. The signal received by LISA is then $h(t) = \onehalf[F^+h_+ + F^{\times}h_{\times}].$ with:
\begin{equation}
F^+h_+ + F^{\times}h_{\times} = [(p_ap_b-q_aq_b)h_+ + (p_aq_b-q_ap_b)h_{\times}](\ell_1^a\ell_1^b-\ell_2^a\ell_2^b).
\end{equation}
Defining
\begin{eqnarray}
A_n &=& \frac{1}{\sqrt{2}}(2F^{\times}\cos{\vartheta}\cos{2\varphi}+F^+(1+\cos^2{\vartheta})\sin{2\varphi})h^{(n)}_{xy}\\
B_n &=& \frac{-1}{2\sqrt{2}}[(2F^{\times}\cos{\vartheta}\sin{2\varphi}-F^+(1+\cos^2{\vartheta})\cos{\varphi})h^{(n)}_{x-y}\nonumber\\
& & ~~~~~~+F^+(1-\cos^2{\vartheta})h^{(n)}_{x+y}],
\end{eqnarray}
we can write the signal  in a standard phase/amplitude form:
\begin{equation}
h(t) = \sum{\sqrt{A_n^2+B_n^2}\cos{[2\pi nft + \phi_{np}(t) + \phi_{nD}(t)]}}
\end{equation}
where the changing orientation of LISA contributes to a time-dependent phase, $\phi_{np}(t)$, due to the varying sensitivity to each polarization and the orbital motion of LISA about the sun contributes to a time-dependent phase, $\phi_{nD}(t)$, due to the Doppler shift in the received signal.

The angles $\vartheta$ and $\varphi$ are determined by the source position and orientation by:
$\cos{\vartheta} = -\hat{n}\cdot\hat{L}$ and $\tan{\varphi} = (\hat{n}\cdot\hat{b})/(\hat{n}\cdot\hat{a})$.
The polarization phase and Doppler phase are: $\phi_{np} = \tan^{-1}{(-A_n/B_n)}$ and $\phi_{nD} = 2\pi nf(R/c)\sin{\theta_s}\cos{(\phi(t)-\phi_s)}$ with $R = 1 {\rm AU}$, and $(\theta_s,\phi_s)$ the angular position of the source. The center-of-mass trajectory of LISA is $\phi(t) = 2\pi t/T$ with $T = 1 yr$.

A second signal is generated from a linear combination of the responses from all three arms of LISA which is equivalent to the signal from a detector rotated by $45\arcdeg$ in the plane of the initial detector \cite{cutler}. These two signals are used to calculate the signal to noise ratio for eccentric binaries.

The Fourier transform of the signal is $\tilde{h}(f) = \sum{\tilde{h}^{(n)}(f)}$ with $\tilde{h}^{(n)}$ sharply peaked about $nf$. The signal to noise ratio ($\rho$) is found by comparing the strength of the signal at each frequency to the expected instrumental noise at that frequency. The spectral noise density $S_n(f)$ can be considered nearly constant in the small region about $nf$. Using Parseval's theorem and the fact that the amplitude is slowly varying with respect to the harmonic frequency, we have:
\begin{equation}
\rho^2 \simeq \sum_{\rm detectors}{\sum_n{[S_n(nf)]^{-1}\int_0^T{(A_n^2+B_n^2)dt}}}.
\end{equation}

\begin{figure}[h]
\centerline{\epsfig{file=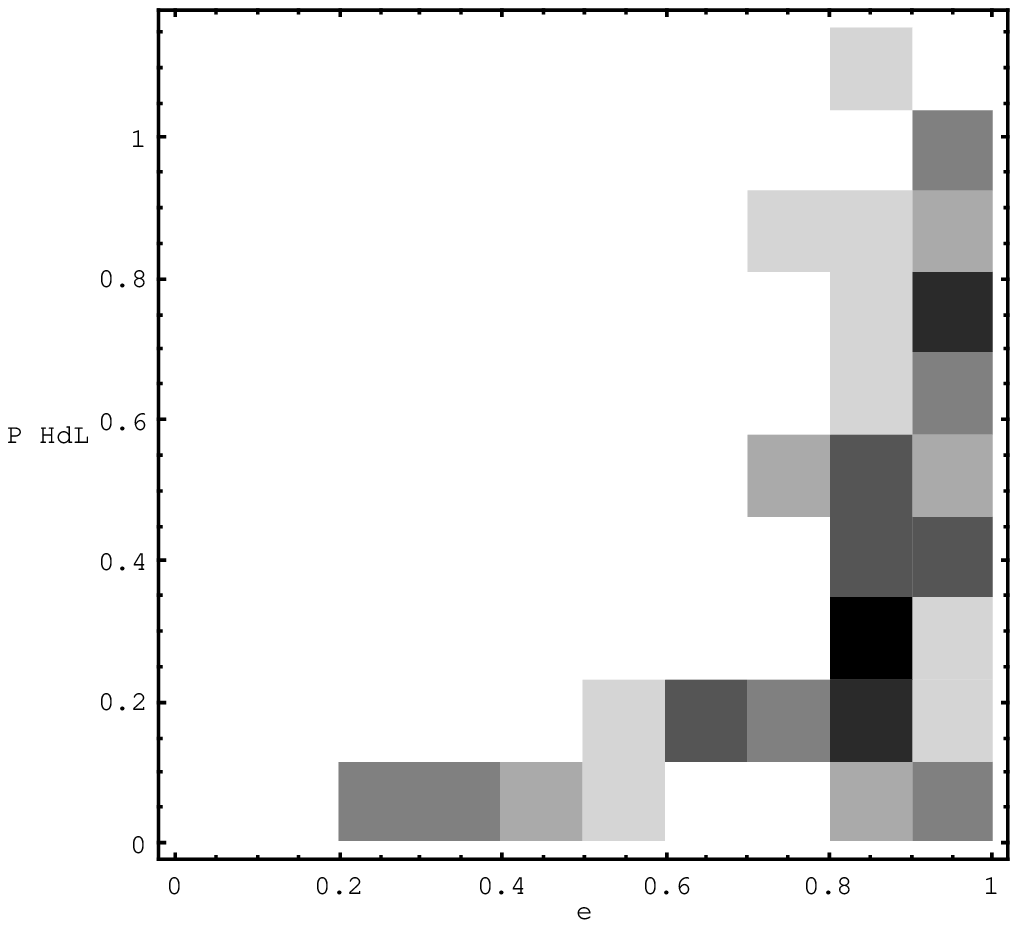,width=2.1in}\hspace{0.4in}\epsfig{file=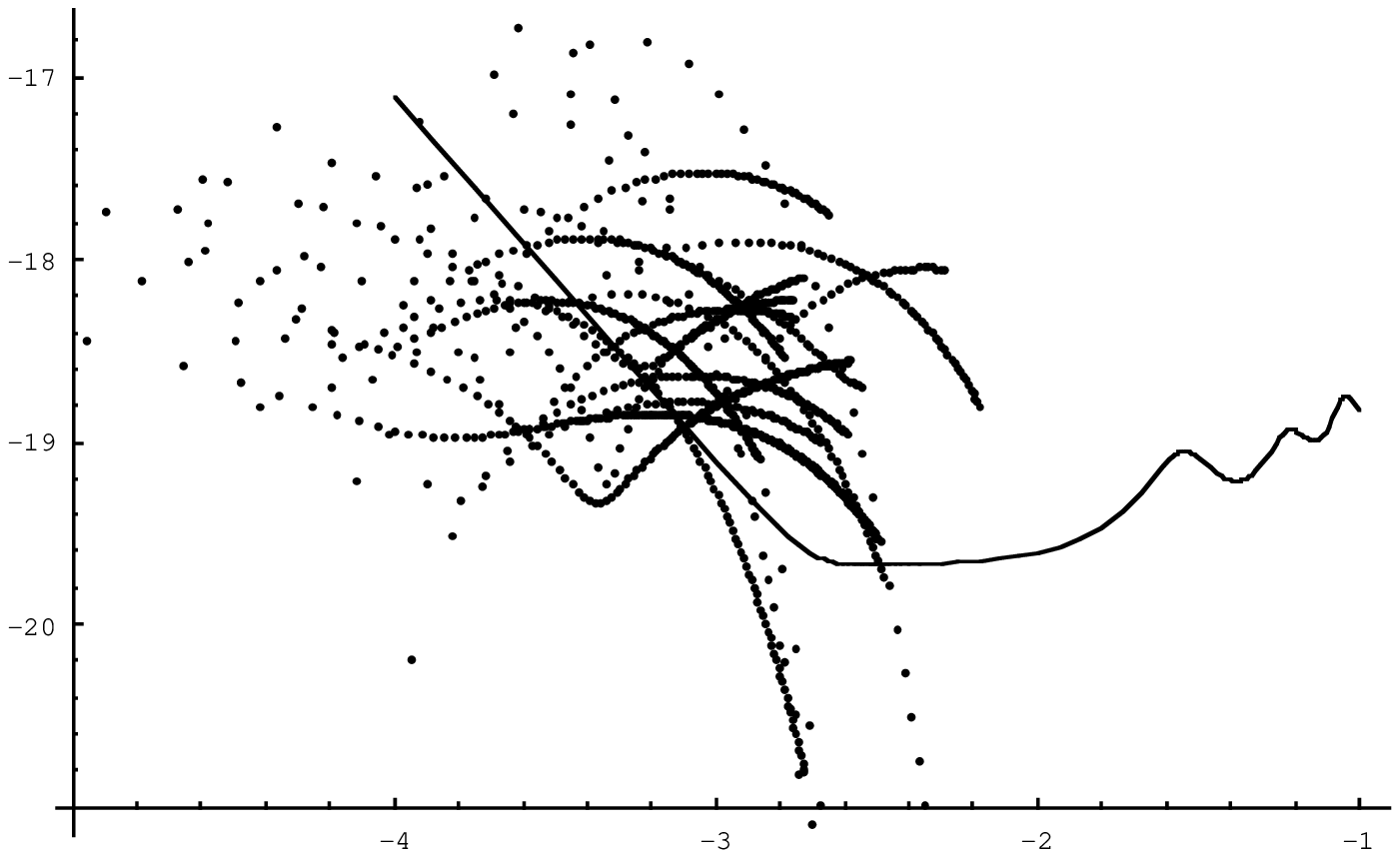,width=3.2in}}
\caption{On the left, a contour plot showing the relative number of detectable binaries for a given range of orbital period (in days) and eccentricity. Darker bins contain more detectable binaries. On the right, the signal strengths from a simulation of 100 globular cluster binaries. Each track represents the harmonics for a single eccentric binary}
\label{contour}
\end{figure}

 We have explored the parameter space of possible eccentric binaries using a Monte Carlo approach. The source position is assigned to be the position of a globualr cluster selected at random from the galactic globular cluster system. The source orientation is assigned with all orientations equally likely. The mass of each component of the binary as well as the eccentricity and orbital frequency are all chosen with equal likelihood from the following ranges: $8 M_{\sun} < M < 18 M_{\sun}$, $0 < e < 1$, and $0.5 {\rm mHz} < f < 0.01 {\rm mHz}$. Out of a total of 300 binaries, 71 were detectable by LISA with $\rho \geq 2$. Most of these had significantly higher values of $\rho$. The number of detectable binaries in each bin of orbital period and eccentricity are shown in Fig~\ref{contour}. The detection of an eccentric binary relies an multiple harmonics appearing in the data stream. The signal strengths from a simulation of 100 globular cluster binaries are also shown in Fig~\ref{contour}.

\end{document}